# Self-Assembly of a halogenated organic molecule on the Si(111) √3×√3-Ag surface

R Liu[1], D. Marchese[1], R. C. Mawhinney[2], & M.C. Gallagher[1†]

1. *Department of Physics, Lakehead University, Thunder Bay, Ontario P7B-5E1, CANADA*

2. *Department of Chemistry, Lakehead University, Thunder Bay, Ontario P7B-5E1, CANADA*

---

[†] Corresponding author: e-mail mark.gallagher@lakeheadu.ca; Tel 1-(807) 343-8110 ext 8226.

## ABSTRACT

We study the self-assembly of halogen-based organic molecules on a passivated silicon surface. The room temperature adsorption of 2,4,6-tris(4-iodophenyl)-1,3,5-triazine (TIPT) on the Si(111)-√3×√3-Ag surface is described. The adsorption is investigated primarily by room-temperature scanning tunneling microscopy (STM) and density-functional theoretical (DFT) calculations. The experimental results is are a dramatic example of how the substrate can influence the structure of the self-assembly. With increasing dose, the TIPT monomers form supramolecular structures defined by a two monomer, $2.07 \pm 0.05$ nm by $1.83 \pm 0.05$ nm rectangular cell. The unit cell is characterized by zig-zag rows of molecules aligned ±13° from the high symmetry directions of the √3-Ag substrate. The 2.07 nm dimension along the zig-zag rows is very similar to self-assembled TIPT networks observed on HOPG, however the 1.83 nm dimension is extended considerably and is commensurate with the √3-Ag substrate. The epitaxial relationship between the overlayer and the substrate, and the commensurate extension of the unit cell indicate significant molecule-substrate interactions. In fact, DFT calculations of free standing TIPT hexamers reveal that increasing the inter row TIPT spacing comes at little energy cost. Experiments also indicate that the formation of supramolecular TIPT domains is extremely sensitive to the quality of the underlying √3-Ag reconstruction. Point defects in the √3-Ag reconstruction ultimately restricts the extent of the observed domains.

**INTRODUCTION**

Molecular self-assembly is a powerful strategy to create two-dimensional (2D) organic nanostructures in which the properties can be engineered by careful choice of monomer architecture. The balance between molecule−molecule and molecule−surface interactions drives the formation of compact 2D nanostructures. Halogen terminated molecules have been demonstrated as attractive candidates for molecular self-assembly given the directionality, and tunability of the possible noncovalent interactions present[1]. It has been shown that halogen⋯halogen or halogen⋯hydrogen based interactions can stabilize two-dimensional (2-d) supramolecular networks. Such networks have been reported on noble metal or HOPG surfaces[2–6].

In contrast to single crystal metal surfaces, very limited work has been done to explore the self-assembly of halogenated organic molecules on silicon. An active molecular layer on a silicon surface could form the basis of a hybrid device and be incorporated into existing Si electronics technology[7,8]. A significant challenge to molecular self-assembly on silicon is the high reactivity of the surface to organic molecules which limits surface diffusion of the adsorbed species[7]. To overcome this reactivity, the surface needs to be passivated to allow for the supramolecular ordering of the halogenated monomers. The self-assembly of halogenated molecules has been reported on the Si(111) $\sqrt{3}\times\sqrt{3}$-B surface[9,10]. The Si(111) $\sqrt{3}\times\sqrt{3}$-Ag surface is also a suitable substrate for molecular self-assembly[11–17], although there are relatively few examples of the self-assembly of halogenated molecules. Liu et al.[16] observed the formation of supramolecular domains of brominated tetrathienoanthracene (TBTTA) molecules on the $\sqrt{3}$-Ag surface and Tsukahara and Yoshinobu[17] observed the compact self assembly of 1,3,5-tris(4-bromophenyl)benzene (TBB) molecules on the same surface.

In this paper, we study the adsorption of the halogenated organic molecule 2,4,6-tris(4-iodophenyl)-1,3,5-triazine (TIPT), onto the Si √3-Ag surface in ultrahigh vacuum (UHV) using room temperature scanning tunneling microscopy (STM). TIPT was chosen because it is observed to form high-quality self-assembled molecular networks (SAMNs) on both HOPG and Au(111) surfaces [2], and 2-d polymers on Cu(111) and Ag(111)[18].

**EXPERIMENTAL AND THEORETICAL METHODS**

All measurements are performed in a single ultra-high vacuum (UHV) system with a base pressure of approximately $2 \times 10^{-10}$ Torr. The Si(111) wafers (n-type, phosphorus doped, ~ 1 Ω·cm) are oriented 1 degree towards $[1\bar{1}0]$[†]. After degassing the samples are flashed to 1200 ºC for several seconds, held at 1060 ºC for 1 min, followed by a 1 min anneal at 850 ºC. This procedure has been shown to ensure high quality 7×7 domains. The efficacy of this process is verified using low energy electron diffraction (LEED), Auger electron spectroscopy (AES), and STM.

The √3 × √3-Ag layer is formed by exposing the silicon surface to a net Ag flux of 1 ML (1 ML is defined as $7.8 \times 10^{14}$ Ag atoms/cm$^2$). The Ag is evaporated from a tantalum basket. Following deposition, the sample is annealed at 580 ºC for 2 min. The resultant surface yields a strong √3 × √3 R30° diffraction pattern as observed in LEED.

---

[†] Siltronix, 74, 160 Archamps, France.

The TIPT molecules are synthesized according to a procedure reported previously[2]. To deposit the TIPT monomers onto the √3 surface, the molecules are sublimed *in-situ* from a boron-nitride crucible onto the √3-Ag surface which is at ambient temperature. Molecular deposition is confirmed using AES (Figure S1).

All STM measurements are performed at room temperature using electrochemically etched W tips. The STM images are obtained using an Omicron MicroSPM[‡] instrument with RHK SPM100[§] control electronics and collected in constant current mode. All bias voltages, $V_s$, represent the sample bias measured with respect to the tip.

Gas-phase DFT calculations are performed using the Gaussian 16 suite of programs[19]. All structures are optimized using a hybrid PBE1PBE functional[20] which combines an exact Hartree-Fock (HF) exchange with a Perdew-Burke-Ernzerhof (PBE) exchange, and uses the PBE correlation functional as formulated within a generalized-gradient-approximation (GGA). We use the Los Alamos National Laboratory Lanl2DZ basis sets[21–24].

A hexameric structure is used to model the supramolecular unit cell. To facilitate cell size variation, starting with an optimized free-standing hexamer, we adjust the cell size by constraining the distances and angles between iodine atoms as indicated in Scheme 1, and maintain $C_{2v}$ symmetry. All other geometric parameters are allowed to relax.

---

[‡] Scienta Omicron GmbH, Taunusstein, Germany.
[§] RHK Technology Inc., Troy MI, USA.

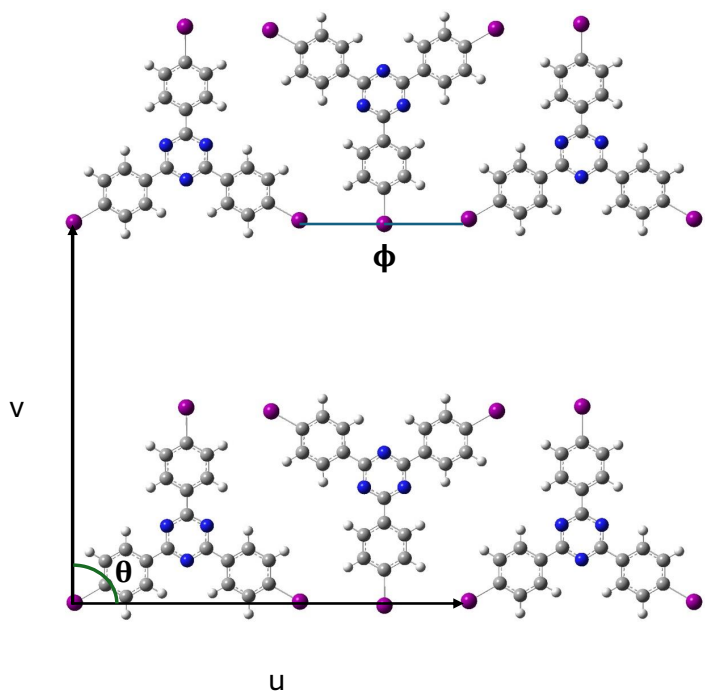

*Scheme 1*: *imposed geometric constraints on the calculated hexamer.*

**RESULTS AND DISCUSSION**

At low coverage the TIPT monomers are quite mobile on the √3-Ag surface (Figure S2). Many STM images exhibit horizontal streaks in the fast scan direction. This apparent 'noise' is attributed to diffusing species which remain trapped below the tip along the fast scan direction. The phenomenon is not uncommon when imaging weakly bound molecular adsorbates at room temperature[25]. In the case of TIPT, rather than seeing the streaks dispersed randomly about the √3 terraces, they tend to cluster and form "fuzzy lines" (Figure S2). The lines suggest that there are positions on the surface where the diffusing species spend relatively more time. We observe that many of these fuzzy lines pass through point defects on the √3 reconstruction and suggests that while point defects are not sufficient to pin the diffusing species, they do lead to increased residence times for monomers in the vicinity. In addition to fuzzy lines, with increasing TIPT exposure we also observe the formation of ordered supramolecular domains (Figure 1).

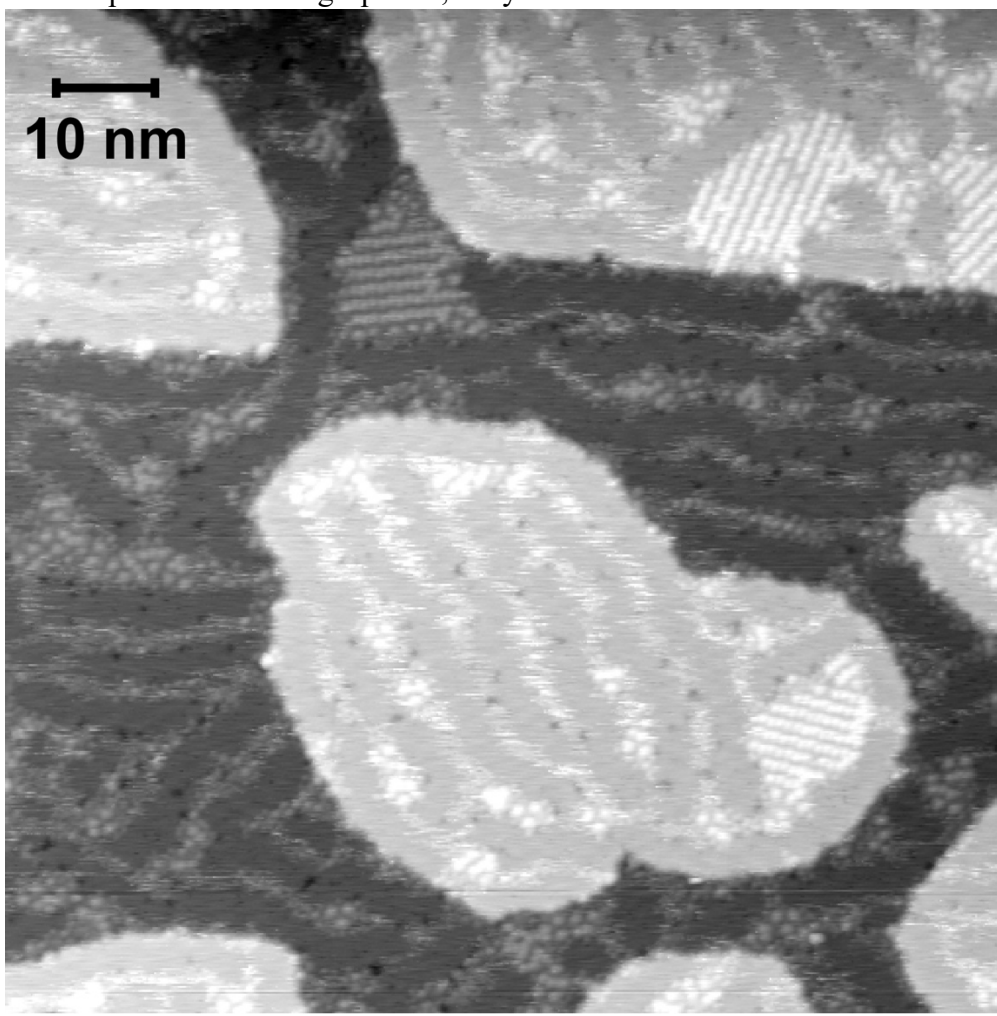

*Figure 1.* STM image of TIPT molecules on the √3-Ag surface ($V_s$ = -1.42 V, $I_t$ = 500 pA). In addition to "fuzzy lines" several supramolecular domains are evident.

On closer inspection the ordered TIPT domains are characterized by a parallel row structure. The rows are "zig-zagged" and the domains can be defined by a rectangular unit cell containing two bright features per cell (Figure 2). The cell dimension along the rows is 2.07 ± 0.05 nm, with an inter-row spacing of 1.83 ± 0.05 nm, and an angle between the two vectors of 90 ± 5 degrees. The molecule-substrate interaction is sufficiently strong to impose a definite epitaxial relationship between the √3-Ag substrate and the TIPT domains. With the ability to simultaneously image both the supramolecular domains and the √3 reconstruction of the substrate we observe that the molecular row direction is rotated by approximately ±13° from the threefold high symmetry directions of the underlying √3-Ag substrate. The offset leads to six possible orientations for supramolecular domains on the surface (four are evident in Figure 1).

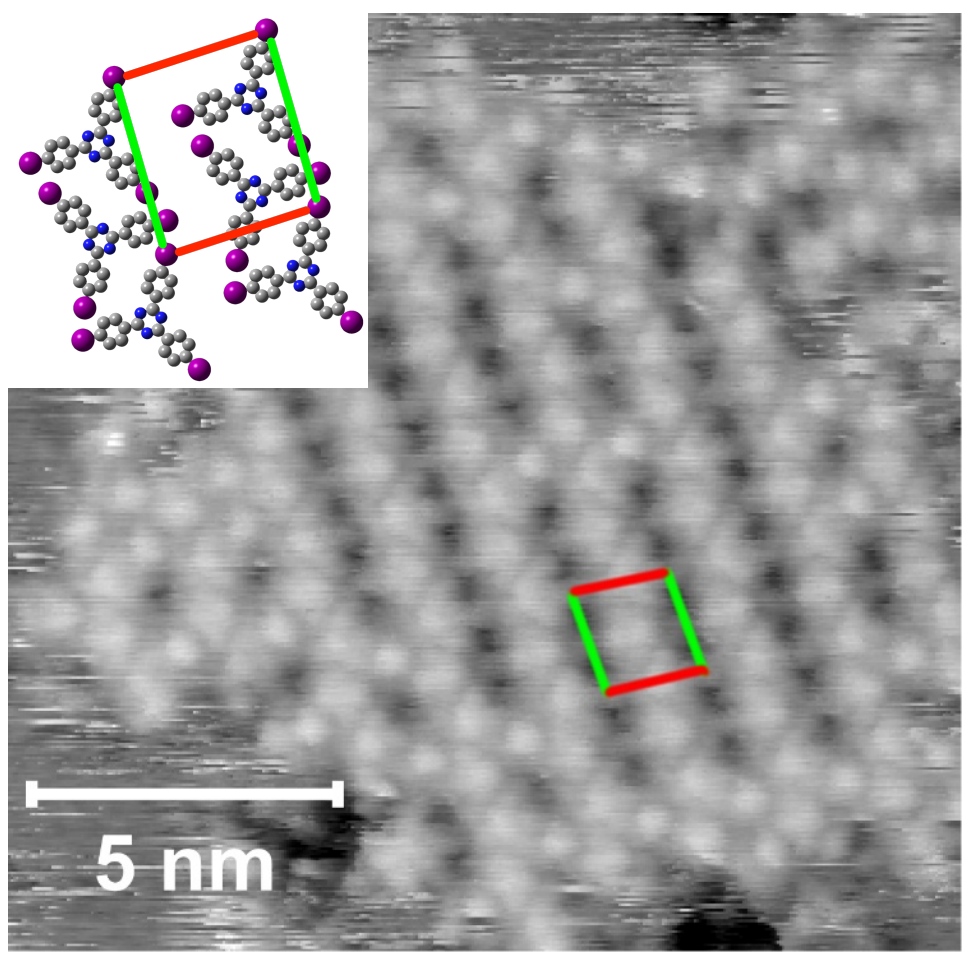

*Figure 2.* STM image of a TIPT supramolecular domain on the √3-Ag surface ($V_s$ = - 1.30 V, $I_t$ = 300 pA). The unit cell is highlighted and a two monomer cell is proposed (see insert).

We believe that the supramolecular regions we observe are domains of intact TIPT monomers. This is based on the high mobility of the monomers at low coverage (Figure 1) and the similarity between the unit cell we observe on the √3-Ag surface and the cell reported previously on HOPG. Gatti et al.[2] investigated the adsorption of TIPT on the HOPG surface at the trichlorobenzene (TCB)-HOPG liquid-solid interface. They found that TIPT forms a SAMN characterized by a two-monomer rectangular unit cell of 2.04 ± 0.09 nm, 1.54 ± 0.09 nm, and an angle of 90° ± 5°. The unit cell dimension of the supramolecular structures we observe along the zig-zagged TIPT rows is the same within error as observed on HOPG: 2.07 ± 0.05 nm compared with 2.04 ± 0.09 nm. In contrast, the inter-row dimensions are quite different: 1.83 ± 0.05 nm on √3-Ag; verses 1.54 ± 0.09 nm on HOPG. In Figure 3 we have superimposed the unit cell of the TIPT overlayer and the √3-lattice of the substrate incorporating the 13° offset observed experimentally (note: exact registry between the molecular overlayer and specific sites on the √3 reconstruction is not known). Expressing the overlayer in matrix form:

$$\begin{pmatrix} \vec{a}_m \\ \vec{b}_m \end{pmatrix} = \begin{pmatrix} 3.4 & 0.8 \\ 1 & 3 \end{pmatrix} \begin{pmatrix} \vec{a}_{\sqrt{3}} \\ \vec{b}_{\sqrt{3}} \end{pmatrix}$$

It is apparent that the 1.83 ± 0.09 nm inter-row spacing is commensurate with the √3 lattice, whereas the 2.04 ± 0.09 nm dimension defining the molecular spacing along the zig-zag rows is incommensurate.

Tsukahara and Yoshinobu[17] reported the self assembly of a related molecule on the Si(111) √3 × √3-Ag surface. TBB is a similar star shape has a benzene core and is Bromine terminated. In contrast to the zig-zag row structure exhibited by TIPT, TBB forms a compact layer with hexagonal symmetry.

The structure of any self-assembled overlayer is determined by the delicate balance between molecule-substrate and molecule-molecule interactions. This is quite evident in the case of star-shaped halogenated molecules like TIPT, which in seemingly similar experiments on different substrates can lead to quite different self-assembly. A comprehensive review of these variations was conducted by Ibenskas and Tornau[26] who attempted to explain the differences in terms pairwise intermolecular interactions. In the case of TIPT, calculations by Gatti et al.[2] and Ibenskas and Tornau[26] indicate that the dominant molecule-molecule interactions are between molecules along the molecular rows. Gatti et al.[2] performed DFT calculations on a free standing 2-d TIPT layer and identified two energy minimized polymorphs with unit cell 1 (2.41 nm × 1.38 nm), and unit cell 1* (2.19 nm × 1.54 nm) exhibiting energies of -0.038 eV/nm$^2$ and -0.072 eV/nm$^2$ respectively. Both configurations position the positively charged tip of the C−I bonds in the direction of the negatively charged belts of neighboring C−I bonds however they concluded that the iodine atoms are too far apart to engage in significant halogen bonding and rather the SAMN is stabilized primarily by I⋯H hydrogen bonding contacts. Ibenskas and Tornau[26] performed free standing DFT calculations on a similar 2,4,6-tris(4-bromophenyl)-1,3,5-triazine (TBPT) layer and concluded that the layer is held together primarily by a double bond between Br atoms and the two H atoms on adjacent monomers, i.e. intra-row interactions dominate.

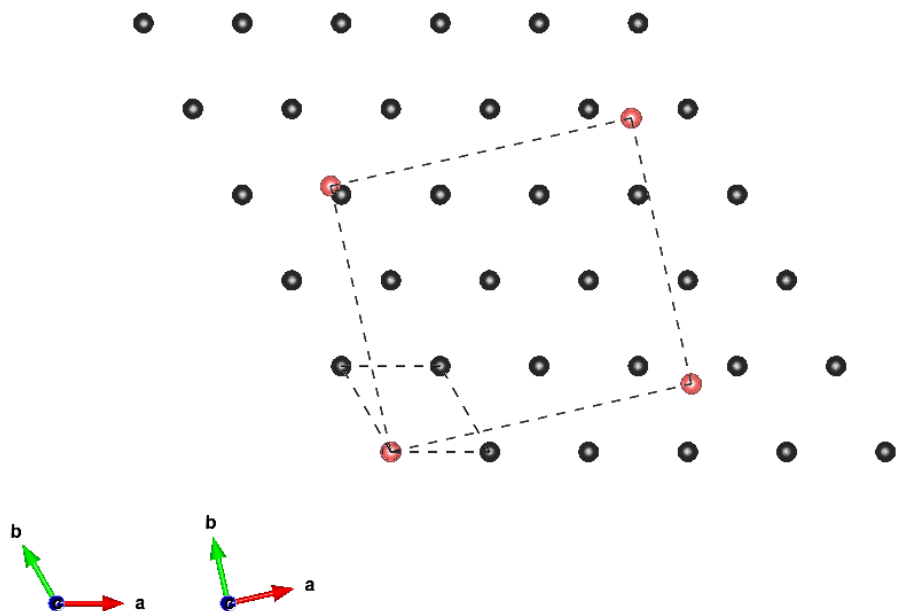

*Figure 3.* *The unit cell of the TIPT overlayer (in red) superimposed on the Si(111) √3 × √3-Ag lattice structure incorporating the 13° offset observed in STM.*

The incommensurate 2.04 ± 0.09 nm dimension along the zig-zag rows of the TIPT layer we observe agrees with both the spacing on HOPG and free standing DFT calculations[2]. We believe the row spacing of the TIPT overlayer is dominated by molecule-molecule interactions along the zig-zag rows, and is seemingly insensitive to the lattice spacing of the substrate. As discussed, calculations by Ibenskas and Tornau[26], indicate that the intermolecular interactions between rows are relatively weak.

To test this hypothesis, we performed DFT calculations on free-standing TIPT hexamers. The hexamers are constructed to define a two-molecule rectangular unit cell.

The molecules are constrained to remain in-plane and are not allowed rotate within the plane. We have calculated the binding energy of a hexamer with dimensions similar to the cell calculated by Gatti et al.[2], in particular u = 2.41 nm × v = 1.39 nm, and $\theta = 90°$ with an angle between adjacent iodine atoms of $\phi = 141°$ (see Scheme 1) and which yields a value of 2.2 kcal/mol. Likewise, we calculated the binding energy of a hexamer with an extended inter-row spacing u = 2.41 nm × v = 1.77 nm, with $\theta = 90°$ and $\phi = 141°$ and found a value of 1.3 kcal/mol. Interestingly if the angle between adjacent iodine atoms, $\phi$, within the hexamer is extended to 176° the binding energy is further reduced further to – 3.5 kcal/mol. Our calculations indicate that the energy difference between the three hexamers tested is small and that the overall energy of the hexamer is reduced by extending the inter-row spacing. This confirms the hypothesis that there is little energy cost to extend the inter-row spacing of the TIPT SAMN to adjust to the √3-Ag lattice.

The extent of the supramolecular domains appears limited by defects in the underlying √3-Ag reconstruction. TIPT domains often terminate or have voids where they encounter defects in the √3 surface layer (Figure 4). It was found when preparing the √3-Ag starting surface, that the defect density is extremely sensitive to the annealing temperature following Ag deposition. Annealing at temperatures above 580 °C led to √3 surfaces with higher defect densities. In fact, when the density of point defects exceeded $5 \times 10^{16}$ m$^{-2}$, no ordered supramolecular TIPT domains were observed.

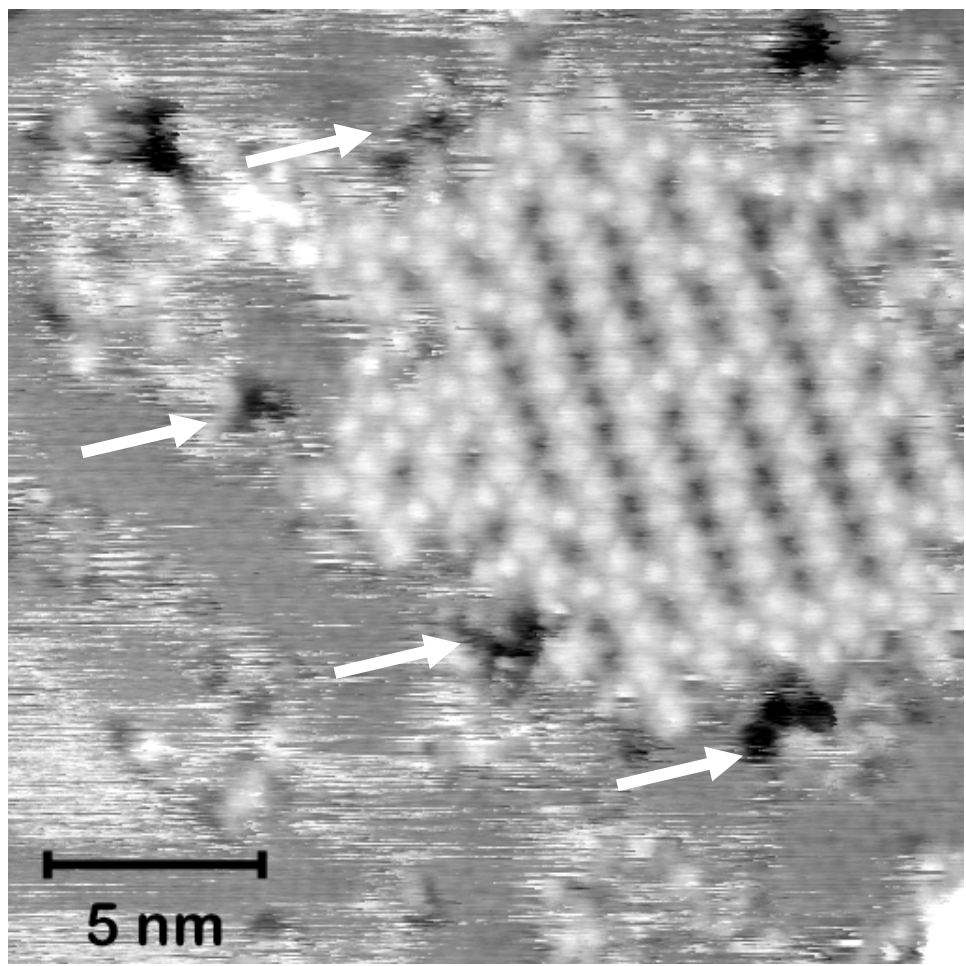

*Figure 4.* STM image of a supramolecular TIPT domain ($V_s$ = -1.42 V, $I_t$ = 490 pA). The extent of the domain is limited by defects in the √3 reconstruction (arrows).

In addition to the inherent defects present after preparation of the √3-Ag layer, the deposition of TIPT molecules at room temperature leads to a further increase in the defect density. Figure 5(a) is an image of a bare Si(111) √3×√3-Ag surface prior to molecular deposition. In comparison Figure 5(b) is the same surface following TIPT deposition. TIPT supramolecular domains are observed, however it is also evident that the number of point defects in the √3 layer has increased. We attribute the increase to dehalogenation of a small fraction of the TIPT molecules, and a resultant destructive interaction between the liberated iodine atoms and the √3-Ag layer. The increased defect density tends to limit the extent of supramolecular domains. With increasing coverage, the surface consists of more ordered supramolecular domains of limited size.

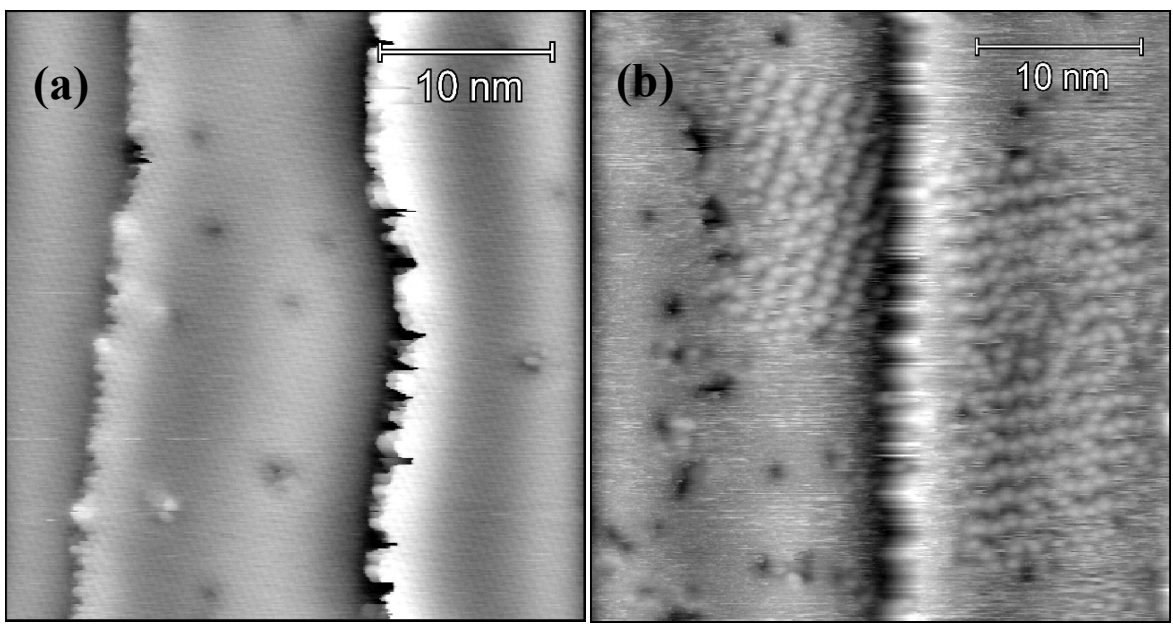

*Figure 5:* *Two STM images taken from the same sample. The bare √3-Ag/Si(111) (a), and the same surface after molecular deposition at room temperature (b).*

**CONCLUSIONS**

We find that the √3-Ag surface provides a high-mobility template for TIPT adsorption at room temperature. At low coverage, intact TIPT monomers readily diffuse to step edges and defects in the √3 overlayer. Many of these images display regularly spaced "fuzzy lines" indicating molecular diffusion at room temperature. At higher coverage, supramolecular domains are formed and defined by a two-monomer rectangular unit cell of 2.07 ± 0.05 nm by 1.83 ± 0.0 nm.

Aspects of the cell observed on the √3-Ag surface are like the TIPT cell observed by Gatti et al.[2] at the TCB-HOPG liquid-solid interface and provides insight into the balance between molecule−molecule and molecule−surface interactions which drive molecular self-assembly. The incommensurate 2.07 ± 0.05 nm dimension along the zig-zag rows is the same as that observed on HOPG and along with DFT calculations suggests that the intra-row spacing is determined primarily by molecule-molecule interactions. In contrast, the commensurate 1.83 ± 0.05 nm inter-row spacing is extended significantly. To investigate the influence of the substate on the inter-row separation we performed DFT calculations of free standing TIPT hexamers. Our results confirm that the energy cost associated with extending the inter-row spacing is relatively low and the commensurate TIPT cell dimension perpendicular to the zag-zag rows is determined by molecule-surface interactions. The how subtle changes can have dramatic effects on self-assembly is also evident in the fact that TIPT and TBB form very different SAMNs on the √3-Ag surface.

The formation of the TIPT supramolecular domains is very sensitive to the quality of the √3 starting surface. Domain size is often limited by point defects in the underlying

√3 reconstruction. In addition, we observe that the number of defects in the √3 layer increases with TIPT coverage. We ascribe the increase to a destructive interaction between silver atoms in the √3 reconstruction and iodine adatoms liberated via the dehalogenation of a small fraction of the TIPT monomers.

**SUPPORTING INFORMATION**

I.  **Auger Electron Spectroscopy (AES)**

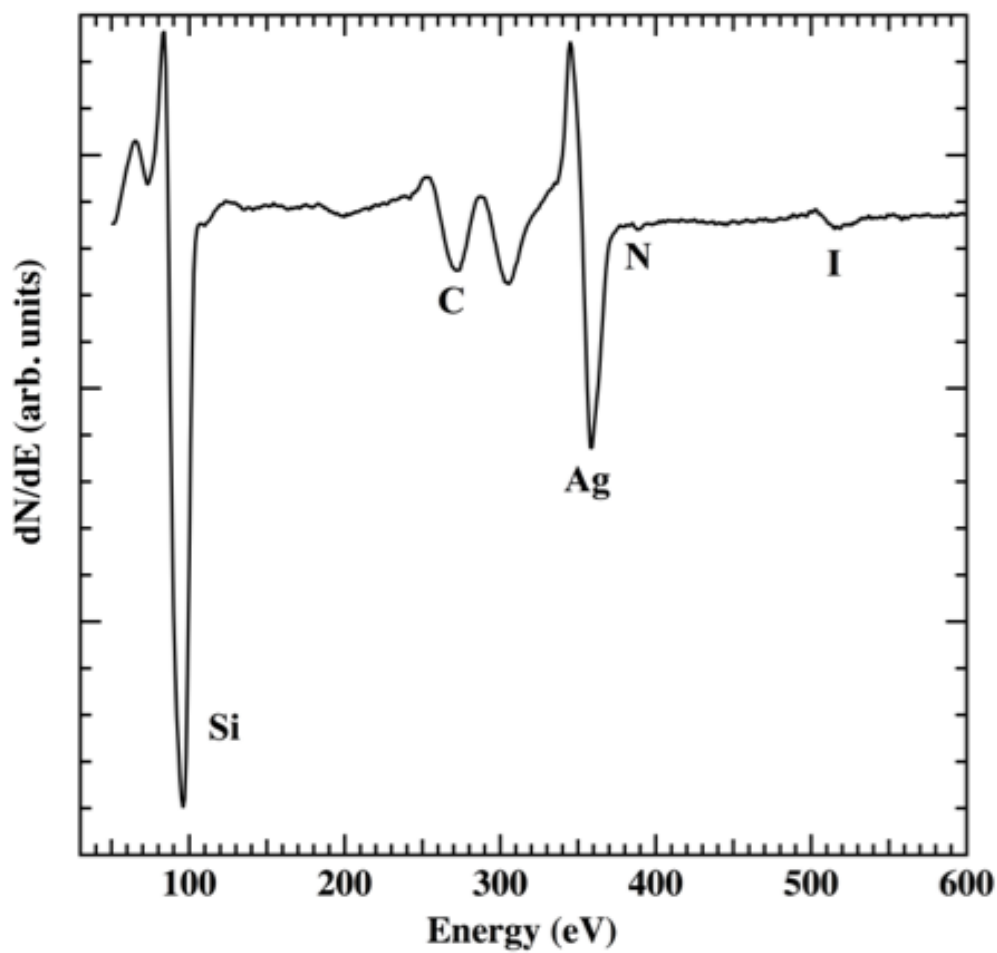

**Figure S1:** An Auger spectrum of TIPT on the Si(111) √3 × √3-Ag surface. The size of the iodine MNN transition at 519 eV relative to silicon LMM transition at 96 eV is used to estimate the molecular coverage.

The AES emission spectra is acquired following excitation by an incident electron beam with a kinetic energy of 3 keV. The spectra reveals an iodine MNN transition at 519 eV, and a carbon KLL transition at 272 eV. The carbon transition is partially masked by a silver (MNN) transition at 351 eV. A crude measure of molecular coverage is made by monitoring the change in amplitude of the iodine peak with respect to the silicon LMM transition at 96 eV. A small nitrogen Auger transition at 389 eV is also observed, however this peak was not tracked due to the low signal strength. Corresponding STM images indicate a molecular coverage close to one monolayer.

## II. Low Coverage Behavior

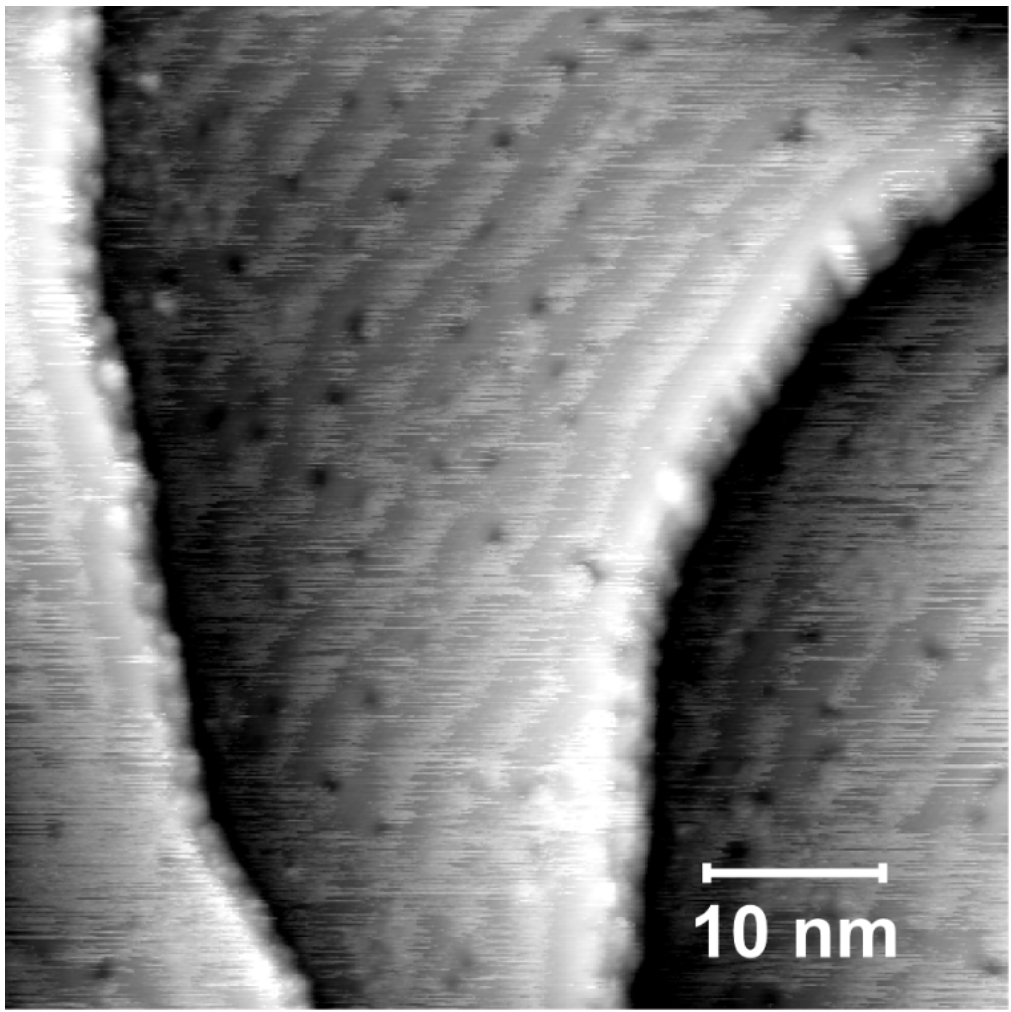

**Figure S2:** An STM image of TIPT on the √3-Ag surface at low coverage. (Vs = -1.22 V, It = 490 pA). The image indicates three terraces separated by atomic steps. Dark spots are point defects in the √3-Ag reconstruction and the. "Fuzzy lines" indicate diffusing species on the surface.

At low coverage the data indicates that the TIPT monomers are quite mobile on the √3-Ag surface at room temperature. With increasing molecular coverage, many STM images exhibit horizontal streaks in the fast scan direction (Figure S2). This apparent ' noise' is attributed to diffusing species which remain trapped below the tip along the fast scan direction. In the case of TIPT, rather than seeing the streaks randomly dispersed about the √3 terraces, they tend to cluster and form "fuzzy lines". The lines would suggest that there are positions on the surface where the diffusing species spend relatively more time. In fact, we observe that many of these fuzzy lines pass through point defects on the √3 reconstruction and suggests that while point defects are not sufficient to pin the diffusing species, they can lead to increased residence times for monomers in the vicinity.

## ACKNOWLEDGMENTS

We would like to thank Professor Dmitrii Perepichka and his group for providing the TIPT molecules. This research was partially supported by grants from the Natural Sciences and Engineering Research Council (NSERC) of Canada.

Page 22 of 23(15) Suzuki, T.; Lutz, T.; Payer, D.; Lin, N.; Tait, S. L.; Costantini, G.; Kern, K. Substrate Effect on Supramolecular Self-Assembly: From Semiconductors to Metals. *Physical chemistry chemical physics : PCCP* **2009**, *11* (30), 6498–6504. https://doi.org/10.1039/b905125b.

(16) Liu, R.; Fu, C.; Perepichka, D. F.; Gallagher, M. C. Supramolecular Structures of Halogenated Oligothiophenes on the Si(111)-√3×√3-Ag Surface. *Surface Science* **2016**, *647*, 51–54. https://doi.org/10.1016/j.susc.2015.12.001.

(17) Tsukahara, N.; Yoshinobu, J. Substrate-Selective Intermolecular Interaction and the Molecular Self-Assemblies: 1,3,5-Tris(4-Bromophenyl)Benzene Molecules on the Ag(111) and Si(111) (√3 × √3)-Ag Surfaces. *Langmuir* **2022**, *38* (29), 8881–8889. https://doi.org/10.1021/acs.langmuir.2c00991.

(18) Galeotti, G. Two-Dimensional Polymer Formation upon Metallic with Low Miller Index, University of Rome Tor Vergatta, 2012.

(19) Frisch, M. J.; Trucks, G. W.; Schlegel, H. B.; Scuseria, G. E.; Robb, M. A.; Cheeseman, J. R.; Scalmani, G.; Barone, V.; Petersson, G. A.; Nakatsuji, H.; Li, X.; Caricato, M.; Marenich, A. V.; Bloino, J.; Janesko, B. G.; Gomperts, R.; Mennucci, B.; Hratchian, H. P.; Ortiz, J. V.; Izmaylov, A. F.; Sonnenberg, J. L.; Williams; Ding, F.; Lipparini, F.; Egidi, F.; Goings, J.; Peng, B.; Petrone, A.; Henderson, T.; Ranasinghe, D.; Zakrzewski, V. G.; Gao, J.; Rega, N.; Zheng, G.; Liang, W.; Hada, M.; Ehara, M.; Toyota, K.; Fukuda, R.; Hasegawa, J.; Ishida, M.; Nakajima, T.; Honda, Y.; Kitao, O.; Nakai, H.; Vreven, T.; Throssell, K.; A, J. M., J.; Peralta, J. E.; Ogliaro, F.; Bearpark, M. J.; Heyd, J. J.; Brothers, E. N.; Kudin, K. N.; Staroverov, V. N.; Keith, T. A.; Kobayashi, R.; Normand, J.; Raghavachari, K.; Rendell, A. P.; Burant, J. C.; Iyengar, S. S.; Tomasi, J.; Cossi, M.; Millam, J. M.; Klene, M.; Adamo, C.; Cammi, R.; Ochterski, J. W.; Martin, R. L.; Morokuma, K.; Farkas, O.; Foresman, J. B.; Fox, D. J. *Gaussian 16 Rev. C.01*; Wallingford, CT, 2016.

(20) Adamo, C.; Barone, V. Toward Reliable Density Functional Methods without Adjustable Parameters: The PBE0 Model. *J. Chem. Phys.* **1999**, *110* (13), 6158–6170. https://doi.org/10.1063/1.478522.

(21) Dunning, Thom. H.; Hay, P. J. Methods of Electronic Structure Theory. **1977**, 1–27. https://doi.org/10.1007/978-1-4757-0887-5_1.

(22) Hay, P. J.; Wadt, W. R. Ab Initio Effective Core Potentials for Molecular Calculations. Potentials for the Transition Metal Atoms Sc to Hg. *J. Chem. Phys.* **1985**, *82* (1), 270–283. https://doi.org/10.1063/1.448799.